\documentclass[13 pt]{article}
\usepackage{amssymb}
\newcommand{\one}[1]{\stackrel{1}{#1}}
\newcommand{\two}[1]{\stackrel{2}{#1}}
\newcommand{\betah}{\hat{\beta}}

\def\vphi{\varphi}
\def\vphib{\overline{\varphi}}
\def\tphi{ {\tilde \phi}}
\def\tvphi{ {\tilde \varphi}}

\def\tvphib{\overline{\tilde \varphi}}
\def\zbar{\overline z}

\def\th{\theta}

\newcommand{\pt}{\hat{p}}
\newcommand{\qt}{\hat{q}}
\newcommand{\beq}{\begin{equation}}
\newcommand{\eeq}{\end{equation}}
\newcommand{\bea}{\begin{eqnarray*}}
\newcommand{\eea}{\end{eqnarray*}}
\newcommand{\beqa}{\begin{eqnarray}}
\newcommand{\eeqa}{\end{eqnarray}}

\newcommand{\lambdah}{{{\hat\lambda}}}
\newcommand{\lambdahb}{{\overline {\hat\lambda}}}

\begin{document}
\newfont{\elevenmib}{cmmib10 scaled\magstep1}%

\newcommand{\preprint}{
            \begin{flushleft}
   \elevenmib Yukawa\, Institute\, Kyoto\\
            \end{flushleft}\vspace{-1.3cm}
            \begin{flushright}\normalsize  \sf
            YITP-03-14\\
           {\tt hep-th/0304120} \\ July 2003
            \end{flushright}}
\newcommand{\Title}[1]{{\baselineskip=26pt \begin{center}
            \Large   \bf #1 \\ \ \\ \end{center}}}
\hspace*{2.13cm}%
\hspace*{1cm}%
\newcommand{\Author}{\begin{center}\large
           P. Baseilhac\footnote{
pascal@yukawa.kyoto-u.ac.jp}\ \ \ \ \mbox{and}\ \ \ \ K.
Koizumi\footnote{kkoizumi@cc.kyoto-su.ac.jp}
\end{center}}
\newcommand{\Address}{{\baselineskip=18pt \begin{center}
           $^1$\it Yukawa Institute for Theoretical Physics\\
     Kyoto University, Kyoto 606-8502, Japan\vspace{0.4mm}\\
          $^2$\it Department of Physics,\\
           Kyoto Sangyo University, Kyoto 603-8555, Japan
      \end{center}}}
\baselineskip=13pt

\preprint
\bigskip

\Title{$N=2$ boundary supersymmetry in integrable models\\
and perturbed boundary conformal field theory}\Author

\vspace{- 0.1mm}
 \Address

\vskip 0.6cm

\centerline{\bf Abstract}\vspace{0.3mm}
Boundary integrable models with $N=2$ supersymmetry are considered. For the simplest boundary $N=2$ superconformal minimal
model with a Chebyshev bulk perturbation
we show explicitly how fermionic boundary degrees of freedom arise naturally in
the boundary perturbation in order to maintain integrability and $N=2$ supersymmetry. A new boundary reflection
matrix is obtained for this model and $N=2$ boundary superalgebra is studied. A factorized scattering theory is
proposed for a $N=2$ supersymmetric extension of the boundary sine-Gordon model with either $(i)$ fermionic or
$(ii)$ bosonic and fermionic boundary degrees of freedom. Exact results are obtained for some quantum impurity
problems: the boundary scaling Lee-Yang model, a massive deformation of the anisotropic Kondo model at the
filling values $g=2/(2n+3)$ and the boundary Ashkin-Teller model.

\vspace{0.1cm} .
\vskip 0.8cm

\vskip -0.6cm

\section{Introduction}
In superstring theory, it has recently been shown that certain backgrounds are associated with two-dimensional
integrable massive field theories on the worldsheet \cite{Malda,tseyt}. For instance, the sine-Gordon model at
 its $N=2$ supersymmetric point \cite{Vaf93} corresponds to the simplest generalization of the pp-wave background
 \cite{Gomez} and the $N=2$ supersymmetric sine-Gordon model \cite{Uematsu} was suggested as a good string background
 \cite{Malda}. Similarly, for open superstrings boundary integrable models with $N=2$ supersymmetry arise naturally
 \cite{Hori,Yama}. Exact results
for such integrable field theories are then desirable.

In statistical physics, several low-dimensional systems around their critical points possess a description in terms
 of two-dimensional integrable boundary  quantum field theories. Among the famous examples, one finds the Kondo model
which describes the s-wave scattering of electrons off a magnetic spin impurity.
Another example corresponds to free fermions on the half-line  which describes the scattering of fermions
off a monopole (Callan-Rubakov effect). Also, minimal models and their extensions (with $N=1,2$ supersymmetry
or $W$-symmetry) arise in quantum impurity problems. Although nonperturbative results for (pure) bulk perturbation
 can be derived using certain quantum group restrictions of  (super)sine-Gordon, Bullough-Dodd or more generally
 affine Toda models, in case of an integrable boundary perturbation such a relation is still an open problem
 which needs further analysis.

As explained in \cite{GZ,Warn}, a boundary perturbation of a massive
integrable quantum field theory will
preserve integrability if the perturbing boundary operator possesses a
representation with respect to the underlying chiral algebra on the
half-plane\,\footnote{This can be shown in first order of conformal
perturbation theory. If operators are very relevant, scaling arguments
can be used to show that local conserved currents exists at all
orders in conformal perturbation theory.} {\it and} provided it belongs to the same representation than
the bulk potential. It is important to mention that \cite{Cardy} the choice of
conformal boundary conditions plays a crucial role as it restricts the
representations of the boundary operators.

Among the simplest known examples, one finds the critical Ising
field theory (with central charge $c=1/2$) perturbed by the energy
operator  in the bulk ($\Phi_{13}$ representation of the Virasoro
algebra) \cite{Zam89}. Due to the argument \cite{GZ} recalled above, it follows
that a boundary operator in the $\Phi_{13}$ representation leads to a
boundary integrable model. Indeed, from the fusion rule of the spin
operator $\sigma\sigma\rightarrow {\mathbb{I}}+\Phi_{13}$ a boundary
integrable model can be constructed as an off-critical Ising model
restricted on the half-line with free conformal boundary conditions
perturbed by a boundary spin operator coupled with one fermionic boundary degree
of freedom \cite{GZ}.

As mentionned in \cite{Warn}, it would be interesting to construct a
 nontrivial boundary integrable perturbation of a critical Ising
field theory on the half-line perturbed by a relevant magnetic operator in
the bulk ($\Phi_{12}$ representation). Indeed, if one starts with
standard boundary conditions (free or fixed), one simply cannot obtain
a boundary operator in the representation of the Virasoro
algebra \cite{Cardy}. Starting from certain boundary conditions
changing operators \cite{Cardy}, it is however expected that one might
be able to construct a well-defined boundary operator that preserves
integrability. But in this case, the underlying boundary conformal field theory (BCFT) with these
special boundary conditions has to be identified, which remains an
open question. In order to solve this problem, it has been pointed out
that introducing new degrees of freedom located at the boundary might
provide a useful tool.

In both previous examples, a perturbed boundary minimal model (critical
Ising field theory) is considered for which the
representations of the Virasoro algebra (labeled by the conformal
weights) are in finite number. Fusion rules then restrict drastically
the number of models that can be constructed\,\footnote{For
boundary conformal field theories with higher
symmetries such as $W$-minimal models \cite{FateevLukyanov}, ${\mathbb Z}_n$
models \cite{FateevZamolodchikov},... the same situation occurs. For a discussion about
the relation between boundary Toda models with imaginary coupling and
perturbed boundary $W$-minimal models, see for instance
\cite{Warn}.}. It is probably one reason why the presence of boundary
degrees of freedom seems to be sometimes necessary.

On the other hand, we may now wonder if boundary degrees of
freedom occur in case of BCFTs with a continuous serie of
representations like the Liouville or, more generally, Toda field
theories perturbed simultaneously by bulk and boundary operators
in the same representation. An interesting example is provided by
the sine-Gordon (SG) field theory on the half-line  without
\cite{Sklyanin,GZ} or with \cite{BasDel,BK} degrees of  freedom at
the boundary. In full generality, we define its Euclidean action
as:
\beqa {\cal A}_{bSG}=\int_{-\infty}^{\infty}dy\int_{-\infty}^{0}dx
\Big(\frac{1}{8\pi}(\partial_\nu\phi)^2-2\mu\cos(\betah\phi)\Big)
+ \int_{-\infty}^{\infty}dy\ \Phi_{pert}^{B}(y)\ + \
{\cal A}_{boundary} \label{actionSG} \eeqa
where the interaction between the SG field and the
boundary terms \ ${\cal E}_\pm(y)$ \ with dimension \ $\sim \mu^{1/2}$
\ reads
\bea \Phi_{pert}^{B}(y) \ =\ {\cal E}_-(y) e^{i\betah\phi(0,y)/2}
+ {\cal E}_+(y) e^{-i\betah\phi(0,y)/2}\ . \eea
Here we used the notation $\betah=\beta/\sqrt{4\pi}$ and introduced
the UV mass parameter $\mu$. The last term only depends on the
boundary degrees of freedom, and is only considered if necessary.

If there are no
degrees of freedom at the boundary, the terms \ ${\cal
E}_{\pm}(y)\sim Const.$ \ are just $c$-numbers and we set \
 ${\cal A}_{boundary}\equiv 0$ \ in (\ref{actionSG}). This two-parameter
family\,\footnote{Only boundary parameters are mentionned here. The
bulk mass is sometimes considered as a third (bulk) parameter.} of boundary
integrable models have
been studied in details since many years, for which exact results
\cite{GZ,Gho94,Fri95,Fen94,Sal95,Sko95,Lec95,Fat97,
Dor99,Cor98,Cor99,Cor00,Dor00,Baj02,Mat} have been confirmed using various methods. This model can be
 considered as an integrable
perturbation of the boundary Liouville field theory as suggested in
\cite{FZZ}. Using vertex
operators representations \cite{FZZ,dots,Schulze,Kawai}, it is not difficult to
see in (\ref{actionSG}) that the boundary perturbing operators are
indeed in the same representation than the
bulk ones (for identical bulk and boundary background charges). Without boundary degrees of freedom,
 the UV limit of the boundary SG
model defined in (\ref{actionSG}) can be identified either with the Gaussian field with Neumann boundary
conditions or with the boundary Liouville field theory as considered in \cite{FZZ}.

For certain degrees of freedom at the boundary and generic values of the coupling $\betah$, the classical
Hamiltonian  associated with the model (\ref{actionSG}) above is
integrable \cite{BasDel}. At quantum level, integrability is
preserved \cite{BK} for
\beqa {\cal E}_+(y)=\pm 2e_{UV}\frac{1-\betah^2}{\betah^2}\cosh\pt(y) \ \ \ \ \
\mbox{and}\ \ \ \ \
{\cal E}_-(y)=\pm 2e_{UV}\frac{1-\betah^2}{\betah^2}\cosh\qt(y) \label{bop}
\eeqa
with \ $\big[\pt(\pm \infty),\qt(\pm \infty)
\big]=\alpha\ mod\ (4i \pi)$ \ and \ $e_{UV}$ \ is a boundary mass parameter. Depending on each sign
in (\ref{bop}), the boundary quantization length \  $\alpha/i$
\ associated with the boundary quantum mechanical system is fixed to
\beqa \alpha=i4\pi\frac{(\betah^2-1)}{\betah^2}
\quad (+) \ \ \ \ \ \ \mbox{or}\ \ \ \ \ \ \alpha=i
2\pi\frac{(\betah^2-2)}{\betah^2} \quad (-)\ .\label{relat} \eeqa
Introducing a ``dynamical'' extension \cite{BK} of the Cartan
subalgebra (identified with the topological charge in the SG
model) and using the boundary quantum group structure discovered
in \cite{coidnepo} \footnote{This structure has been further
studied in \cite{delmac,ba,del2} for which new reflection matrices
associated with non-dynamical boundary have been obtained.}
(coideal subalgebra of $U_q(\widehat{sl_2})$), we obtained the
soliton/antisoliton and breathers boundary reflection matrices. In
particular, for generic values of the coupling $\betah$, we found
that the bulk and the boundary masses are locked
together\,\footnote{Up to a canonical transformation of the
boundary degrees of freedom, the UV parameter $e_{UV}$ is fixed by
the boundary non-local conserved charges algebraic structure in
terms of the bulk UV mass parameter $\mu$. Although not shown
explictly, Warner predicted such phenomena \cite{Warn}.}. We refer
the reader to \cite{BK} for more details.

For generic values of the coupling $\betah$, this model deserves
certainly some interest, but taking some special values provides interesting new insights as we
are going to show. Indeed, for \ $\betah^2=2p/p'$ \ with \ $0<p<p'$ \ integers the bulk SG model
is known to be related with the $\Phi_{13}$ bulk perturbation of minimal models. Furthermore, at
 the special point $\betah^2=4/3$ (repulsive regime) it describes an integrable bulk perturbation
  (Chebyshev type) of $N=2$ superconformal minimal model. The SG model with boundary degrees of freedom
   introduced and studied in \cite{BK} being one possible boundary integrable
   perturbation\,\footnote{There, we did not discuss integrable massless models
   with boundary degrees of freedom considered for instance in
   \cite{baz1,baz2}.} of the
   bulk SG model, it is important to analyse in details these special
points.

In this paper we will focus on $\betah^2=4/(2n+3)$, \ $n\in {\mathbb N}$ in (\ref{actionSG}) for several
 reasons. First, it will provide new examples of perturbed conformal field theories with boundary degrees
 of freedom for which the factorized scattering theory is proposed.
Secondly, for generic values of the coupling $\betah$ in the model
(\ref{actionSG}), free parameters can be introduced in the
boundary operators ${\cal E}_\pm(y)$ through a canonical
transformation of $\pt(y),\qt(y)$. Then, it is important to see if
different boundary operators ${\cal E}_\pm(y)$ preserving
integability can be constructed at these special points for which
the explicit dependence in terms of the free parameters might be
changed. Third, for $n=0$ a massive $N=2$ boundary supersymmetry
algebra appears explicitly.

The paper is organized as follows. In the next section, we exhibit
an exact relation between  the simplest
 bulk-boundary perturbed $N=2$ superconformal minimal model and the sine-Gordon model at special point
 $\betah=4/3$ with fermionic boundary degrees of freedom. It shows that, similarly to the
critical Ising model in a boundary magnetic field, boundary degrees of
freedom appear naturally in this model.
The boundary supersymmetric charges are explicitly constructed from the Lagrangean and differ from the
ones proposed in \cite{Warn,Nepo2}.
Although their expressions are similar to the ones in \cite{Nepo2} (which may probably be considered as
``effective'' supersymmetry charges in terms of the boundary structure) the main difference here is the
fermionic boundary algebra
that appears associated with the fermion number operator.
These supersymmetry charges are used to construct a new boundary reflection matrix with $N=2$ supersymmetry
 which possesses two free parameters and is shown to satisfy the boundary Yang-Baxter equations. The
  corresponding massive $N=2$ boundary superalgebra is constructed and the boundary free energy is proposed.
In section 3, a boundary scattering associated with a $N=2$ supersymmetric version of the boundary sine-Gordon
 model with purely fermionic (proposed in \cite{Nepo1}) or a mixing of fermionic and bosonic boundary degrees
 of freedom is proposed. In any case, the supersymmetric part is different from the one suggested in \cite{Warn,Nepo2}.
   In section 4, we present applications of our results to quantum impurity problems. The boundary scaling Lee-Yang
   model is revisited in light of our results. The anisotropic Kondo model is shown to admit a simple massive integrable
    extension at special points $g=2/(2n+3)$ with $n\in\mathbb{N}$, for which scattering amplitudes are proposed.
    Finally, a boundary version of the Ashkin-Teller model is briefly presented.

\section{$N=2$ supersymmetry in the boundary sine-Gordon model}
Among the class of relevant perturbation of $N=2$ superconformal
minimal model with central charge $c=3\ell/\ell+2$, the ones corresponding
to integrable $N=2$ supersymmetric field theories have been studied in
details \cite{9}. Three different types of relevant bulk perturbations can be
considered. Here, we focus on the model associated with a perturbation
of the form (the so-called Chebyshev perturbation):
\beqa
\Phi_{pert}^{bulk}(z,\zbar)= \lambdah \
G_{-\frac{1}{2}}^-{\overline{G_{-\frac{1}{2}}^-}}\Phi_k^+(z,\zbar)\ \  +
\ \ \lambdahb \
G_{-\frac{1}{2}}^+{\overline{G_{-\frac{1}{2}}^+}}\Phi_k^-(z,\zbar)\ ,\label{pertbulk}
\eeqa
where $\lambdah$ is a complex parameter that characterizes the
strength of the perturbation. The chiral primary fields $\Phi_k^\pm$
possess conformal dimensions \ $\Delta_k={\overline \Delta_k}=k/2(\ell+2)$ \
and \ $G_{-\frac{1}{2}}^\pm$ $({\overline{G_{-\frac{1}{2}}^-}})$ \ are
$N=2$ supersymmetry generators. Actually, this perturbation is known to be
the $N=2$ superconformal analogue of the energy perturbation
of the ordinary critical Ising model. Following the case of the
off-critical Ising model in a boundary magnetic field \cite{GZ} and in order
to maintain supersymmetry (see details in \cite{Warn}), it is then
natural to consider the model with the bulk perturbation
(\ref{pertbulk}) on the half-line and a boundary perturbation of the
form \cite{Warn}:
\beqa
\Phi_{pert}^{boundary}(y)= \mu_B e^{-i\phi_0/{\sqrt 3}}\ \nu^{\bold{uv}}_-(y)\
G_{-\frac{1}{2}}^-{\hat{\Phi}_k}^+(y)\ \
+\ \ \mu_B e^{i\phi_0/{\sqrt 3}} \ \nu^{\bold{uv}}_+(y)\ G_{-\frac{1}{2}}^+{\hat{\Phi}_k}^-(y)
\ .\label{pertboundary}
\eeqa
Here ${G_{-\frac{1}{2}}^\mp}{\hat\Phi_k}^\pm(y)$ are boundary operators in
the same $N=2$ superconformal representation as the bulk operators, $\phi_0$ is a phase and $\mu_B$ is
the boundary mass scale. It
is important to stress that the resulting model is neither expected to be
integrable nor $N=2$ supersymmetric for any boundary degrees of freedom $\nu^{\bold{uv}}_\pm(y)$. Also, both requirement
must be simultaneously consistent. This can be checked from the
structure of the Lagrangean, but also in the scattering amplitude. To
proceed further, let us consider a concrete example: the sine-Gordon model
at $\betah^2=4/3$.
%
%
%

In the bulk and for $\ell=1$ the simplest $N=2$ superconformal minimal model possesses a Lagrangean representation
in terms of a single free boson $\tphi(x,y)$ compactified on the ``supersymmetric'' radius. In terms
of its holomorphic and antiholomorphic parts it reads \ $\tphi(x,y)=\tvphi(z)+\tvphib(\zbar)$ \ with $z=y+ix$ and
$\zbar=y-ix$.  If
we denote the expectation value over the Fock vacuum space of massless fields
 $\langle ... \rangle_0$, then the holomorphic/antiholomorphic components  are normalized such that
\bea \langle{\tvphi}(z){{\tvphi}}(w)\rangle_0=-\ln(z-w),\quad
\langle{\tvphib}({\bar z}){\tvphib}({\bar w})
\rangle_0=-\ln({\bar z}-{\bar w}),\quad
\langle{\tvphi}(z){\tvphib}({\bar w})
\rangle_0=0\ .\label{correlbulk}
\eea
The $N=2$ supersymmetric generators correspond to the stress-energy
tensor which ensures conformal invariance, the $U(1)$ current and the two supersymmetry generators. Their holomorphic
parts read, respectively,
\beqa
T(z)=-\frac{1}{2}(\partial_z\tvphi)^2\ ,\qquad \qquad
J(z)=\frac{i}{\sqrt 3}\partial_z\tvphi\ \qquad \mbox{and}\qquad \qquad
G^{\pm}(z)=\exp({\pm i{\sqrt 3}\tvphi(z)})\ .\label{defi}
\eeqa
The order parameter and its conjugate are defined as
\ $\Phi^{\pm}_{k=1}(z,\zbar)=\exp({\pm\frac{i}{\sqrt 3}(\tvphi(z)+\tvphib(\zbar)}))$ \ .
 Using the definitions (\ref{defi}) above,
it follows that the interacting terms in the bulk perturbation
(\ref{pertbulk}) become
\beqa
G_{-\frac{1}{2}}^\pm{\overline{G_{-\frac{1}{2}}^\pm}}
\Phi_k^\mp(z,\zbar)=\exp({\pm\frac{2i}{\sqrt 3}(\tvphi(z)+\tvphib(\zbar)}))\ .
\eeqa

On the half-line, one can use the method of mirror images for Neumann
boundary conditions \ $\partial_x\tphi(x,y)|_{x=0}=0$ \ i.e. \ one defines \  $\phi(x,y)=\tphi(x,y)+\tphi(-x,y)$.
In terms of its holomorphic and antiholomorphic part restricted on the half-line it gives
\beqa
\varphi(z)= \tvphi(z) + \tvphib(z) \qquad \qquad \mbox{and} \qquad \qquad
{\overline\varphi}(z)= \tvphib(\zbar) + \tvphi(\zbar)\ .
\eeqa
from which one deduces the two-point functions in the BCFT with Neumann
boundary conditions
\bea \langle{\varphi}(z){{\varphi}}(w)\rangle_0=-2\ln(z-w),\quad
\langle{{\bar\varphi}}({\bar z}){{\bar\varphi}}({\bar
w})\rangle_0=-2\ln({\bar z}-{\bar w}),\quad
\langle{\varphi}(z){{\bar\varphi}}({\bar
w})\rangle_0=-2\ln(z-{\bar w})\ .\label{correl} \eea
In particular, from these definitions\,\footnote{One has \ $\varphi(z)|_{x=0}=\vphib(\zbar)|_{x=0}$ \ which gives
 \ $\vphi(z)|_{x=0}=\phi(0,y)/2$.} and using the supersymmetric generators restricted on the half-line
  \ $G^{\pm}(z)=\exp({\pm i{\sqrt 3}
\vphi(z)})$ \ the boundary perturbation becomes
\beqa
G_{-\frac{1}{2}}^\pm\Phi_k^\mp(y)\equiv\exp({\mp\frac{i}{\sqrt 3}
(\vphi(z)+\vphib(\zbar)}))|_{x=0}=\exp({\mp\frac{i}{\sqrt 3}
(\phi(0,y)}))\ .
\eeqa
It follows that the action associated with the simplest $N=2$ boundary superconformal minimal model corresponds
to the boundary sine-Gordon model
(\ref{actionSG}) at the special point $\betah^2=4/3$ with the substitutions
\beqa
\mu\  \ \longrightarrow\  \ -{\hat\lambda}\ \ =\ \ -{\overline{\hat\lambda}}\qquad \quad \  \mbox{and}
 \qquad \quad {\cal E}_{\pm}(y)\  \ \longrightarrow
 \ \ \mu_Be^{\pm i\phi_0/{\sqrt 3}}\nu^{\bold{uv}}_{\pm}(y)\ .\label{replace}
\eeqa

At the supersymmetric point, the bulk SG model is in the repulsive
regime (no breathers): the spectrum consists of one soliton and one antisoliton.
In the following, they are denoted \ $|u(\theta)\rangle$ \ and \ $|v(\theta)\rangle$ \
with fermion number \ $+1/2$ \ and \ $-1/2$ \ , respectively and form a two dimensional supermultiplet
\cite{BernLecl,Uematsu,FendleyInt}. The corresponding bulk $S$-matrix is known for a long time $\cite{ZZ}$.
For general values of the coupling in the SG model, four nonlocal charges with fractional spin and one topological
 charge generating the $U_q({\widehat {sl_2}})$ affine quantum envelopping algebra exist \cite{BernLecl}.
However, at $\betah^2=4/3$ they become local (spin $1/2$) and
together with the fermion number operator they generate the
massive superalgebra. If $\theta$ denotes the particle rapidity
and $M$ its mass, in the supermultiplet representation
$\pi_\theta$ they act as
\beqa
&&\pi_\theta({{\cal Q}}_{+})|d(\theta)\rangle=
{\sqrt {2M}}e^{\theta/2}|u(\theta)\rangle\
, \qquad
\pi_\theta({\overline {\cal Q}}_{-})|u(\theta)\rangle=
{\sqrt {2M}}e^{-\theta/2}|d(\theta)\rangle\ ,\\
&&\pi_\theta({{\cal Q}}_{-})|u(\theta)\rangle=
{\sqrt {2M}}e^{\theta/2}|d(\theta)\rangle\
, \qquad
\pi_\theta({\overline {\cal Q}}_{+})|d(\theta)\rangle=
{\sqrt {2M}}e^{-\theta/2}|u(\theta)\rangle\ \label{rep}
\nonumber
\eeqa
and
\beqa
F|u(\theta)\rangle=\frac{1}{2}|u(\theta)\rangle \ , \qquad \qquad F|d(\theta)\rangle=-\frac{1}{2}|d(\theta)\rangle \ .
\eeqa
Using this representation, the massive superalgebra can be constructed \cite{WittenOlive,Uematsu,FendleyInt}
 in terms of the Hamiltonian $H$, momentum $P$ and particle number ${\cal N}$ with eigenvalues on one-particle
 states $E=M\cosh(\theta)$,
$P=M\sinh(\theta)$ and 1 or 0, respectively. Using the definition of the coproduct \cite{Uematsu,FendleyInt}
which acts on multiparticle states one
can check that the bulk $S$-matrix commutes with these supercharges and the fermion number operator.

If we now consider the SG model on the half-line at the special value
$\betah^2=4/3$ as defined above (\ref{actionSG}), using the results of
\cite{BK} together with a scale transformation it follows that the boundary supercharges
\beqa
{\hat {\cal Q}}_{\pm}={\cal Q}_{\pm}+{\bar {\cal Q}}_{\mp}+{\hat\nu}^{\bold{ir}}_{\pm}(-1)^{F}\ \label{bsusy}
 \qquad \quad \mbox{with} \qquad \quad {\hat\nu}^{\bold{ir}}_{\pm}=\mp \frac{2{\sqrt M}}{k}{\nu}^{\bold{ir}}_{\pm}
 \label{susycharges}
\eeqa
 are conserved\,\footnote{To do that, one uses the local expression of the supercharges in terms of the field and
 derives their conservation in first order of conformal perturbation theory. Using scaling arguments, it can be shown
  that this is true at all orders too, which allows to define them asymptotically i.e. in the IR limit.}. Here $k$
  denotes a real parameter.
 To relate the IR and UV data (using the relation in the UV of the sine-Gordon non-local charges and the supercharges),
  asymptotically  we have assumed the boundary conditions for the operators
 $\nu^{\bold{uv}}_{\pm}(y)$ at both ``ends'' of the boundary:
\beqa \nu^{\bold{uv}}_{+}(y=\pm\infty)\sim\nu^{\bold{ir}}_{+}\qquad \qquad \mbox{and}\qquad
 \qquad \nu^{\bold{uv}}_{-}(y=\pm\infty)\sim\nu^{\bold{ir}}_{-}\ .\label{opcond}\eeqa
Similarly to the boundary SG model with  boundary degrees of freedom, the asymptotic boundary operators
 $\nu^{\bold{ir}}_{\pm}$ commute with the
massive (bulk) $N=2$ superalgebra. Thus we have
\beqa \pi_\theta({\hat {\cal Q}}_{\pm})^u_u=-i{\hat\nu}^{\bold{ir}}_\pm \
, \qquad \pi_\theta({\hat {\cal Q}}_{\pm})^u_d={\sqrt {2M}} e^{\pm\theta/2}\ ,
\qquad \pi_\theta({\hat {\cal Q}}_{\pm})^d_u={\sqrt {2M}} e^{\mp\theta/2} \
,\qquad \pi_\theta({\hat {\cal Q}}_{\pm})^d_d= i{\hat\nu}^{\bold{ir}}_\pm\ \nonumber \eeqa
If we impose $N=2$ boundary supersymmetry in the quantum model, the supercharges (\ref{bsusy}) must commute
 with the boundary reflection matrix. Then, we are looking for the ``minimal'' solution of the dynamical
 extension of the intertwining equation considered in \cite{Warn,Nepo2}:
\beqa {K_{\bold {susy}}}^\delta_\nu(\theta){\pi_\theta({\hat
{\cal Q}}_{\pm})}^\nu_\zeta={\pi_{-\theta}({\hat
{\cal Q}}_{\pm})}^\delta_\nu{K_{\bold {susy}}}^\nu_\zeta(\theta) \label{inter}\eeqa
where indices \ $\{\delta,\nu,\zeta\}\in\{u,d\}$ \ refer to two-dimensional supermultiplet representation
 on
asymptotic soliton states \ $|u(\theta)\rangle$, $|d(\theta)\rangle$. Then,
using this representation the solution $K_{\bold {susy}}(\theta)$ is written as
a $2\times 2$ matrix with entries expressed in terms of the
boundary operators. In the non-dynamical case \cite{GZ,Warn}, the entries are
just analytic functions of $\theta$ as $\nu^{\bold{ir}}_\pm$ are
$c-$numbers. Let us define
\beqa {K_{\bold {susy}}}^u_u(\theta)=A(\theta)\ ,\qquad \qquad
{K_{\bold {susy}}}^u_d(\theta)=B(\theta)\ ,\nonumber \\
{K_{\bold {susy}}}^d_u(\theta)=D(\theta)\ ,\qquad \qquad
{K_{\bold {susy}}}^d_d(\theta)=E(\theta)\ .\label{def}\eeqa
After some calculations, we find that the entries of the minimal
solution ${K_{\bold {susy}}}(\theta)$ of the intertwining equation
(\ref{inter}) takes the following form:
\beqa
&&A(\theta)=\Big(\ e^{\theta/2}\nu^{\bold{ir}}_+  \ \ +  \ \ e^{-\theta/2}\nu^{\bold{ir}}_-\ \Big)/k{\sqrt 2} \  ,\qquad \
B(\theta)=\Big(\ -i\sinh(\theta)\ \  +  \ \ \frac{i(\alpha+1)}{k^2}\ [\nu^{\bold{ir}}_+,\nu^{\bold{ir}}_-]\ \Big)/2\ ,\nonumber \\
&&E(\theta)=\Big(\ e^{\theta/2}\nu^{\bold{ir}}_- \  \ +\  \  e^{-\theta/2}\nu^{\bold{ir}}_+\ \Big)/k{\sqrt 2}\ ,\ \qquad
 D(\theta)=\Big(\ -i\sinh(\theta)\ \  + \ \ \frac{i(\alpha-1)}{k^2}\ [\nu^{\bold{ir}}_+,\nu^{\bold{ir}}_-]\ \Big)/2\ ,\label{entrees}
\eeqa
where the boundary operators satisfy
\beqa
\{\nu^{\bold{ir}}_\pm,[\nu^{\bold{ir}}_+,\nu^{\bold{ir}}_-]\}=0 \qquad \quad  \mbox{and} \qquad
\quad [({\nu^{\bold{ir}}_\pm})^2,\nu^{\bold{ir}}_+]=0 \ .\label{condop}
\eeqa
Here $\alpha$ is a real parameter.
One obvious realization is $\nu^{\bold{ir}}_{\pm}\equiv e^{\pm i\xi}$ in which case we recover the
Ghoshal-Zamolodchikov-DeVega-Gonzalez-Ruiz minimal solution
\cite{GZ,DeVega}. An other solution to ({\ref{condop}) is to introduce a dimensionless complex fermion
at the boundary ${\bold b}(y)$, ${\bold b}^\dagger(y)$ \ such that using (\ref{opcond}), asymptotically it satisfies
\beqa
\{{\bold b},{\bold b}\}=\{{\bold b}^\dagger,{\bold b}^\dagger\}=0 \qquad \quad \mbox{and} \qquad \quad
\{{\bold b},{\bold b}^\dagger\}=1 \qquad \quad \mbox{with}\qquad {\bold b}(\pm\infty)\equiv {\bold b}\ .\label{fermiondef}
\eeqa
Together with the asymptotics (\ref{opcond}) we choose the representation $\nu^{\bold{ir}}_{\pm}=\sigma_\pm$ in
(\ref{entrees}), where $\sigma_\pm$ are Pauli matrices which act on the (pure) boundary space of states. In the
Lagrangean (\ref{actionSG}) with (\ref{replace}), it leads to the choice
\beqa
\nu^{\bold{uv}}_{+}(y)=  \ {\bold b}(y) \qquad \quad \mbox{and}\qquad \quad \nu^{\bold{uv}}_{-}(y)= \ {\bold b}^\dagger(y)
  \ .\label{link}
\eeqa
We would like to stress that the minimal part (\ref{def}) with (\ref{entrees}) of the boundary reflection matrix
 associated with the SG model at its $N=2$ supersymmetric point possesses two free parameters \ $\{k,\alpha\}$.
 In full generality, one may choose to introduce a phase $e^{\pm i\rho}$ in front of $\nu^{\bold{ir}}_{\pm}$ in
  (\ref{entrees}). However this phase can be removed using a gauge transformation over the fermionic boundary degrees
   of freedom.

\subsection{Boundary Yang-Baxter equations and boundary reflection matrix}
In the bulk, the perturbed $N=2$ superconformal minimal model (the SG model at $\betah^2=4/3$) is massive and integrable.
 Integrability imposes strong constraints on the system which implies that the $S$-matrix has to satisfy
the quantum Yang-Baxter equations. The entries at $\betah^2=4/3$ read for $u=-i\theta$:
\beqa
{S_{\bold {susy}}}^{\pm\pm}_{\pm\pm}(\theta)\equiv a(\theta)=\cos(u/2)Z(u)\ , \quad {S_{\bold {susy}}}^{\pm\mp}_{\pm\mp}
(\theta)\equiv b(\theta)=\sin(u/2)Z(u)\ , \quad {S_{\bold {susy}}}^{\pm\mp}_{\mp\pm}(\theta)\equiv c(\theta)=Z(u)\
\eeqa
and vanish otherwize. The factor $Z(u)$ ensures unitarity and crossing symmetry of the $S$-matrix. Its exact expression
 can be found in \cite{FendleyInt}.

For boundary integrable field theories, the soliton/antisoliton reflection matrix is constrained by the boundary
Yang-Baxter
equations (the so-called reflection equations). In our case, it reads $R_{\bold {susy}}(\th)$ and obeys
\beqa
\label{qRR} S_{\bold {susy}}(\th-\th'){\one
R_{\bold {susy}}}(\th)S_{\bold {susy}}(\th+\th'){\two
R_{\bold {susy}}}(\th') ={\two
R_{\bold {susy}}}(\th')S_{\bold {susy}}(\th+\th'){\one
R_{\bold {susy}}}(\th)S_{\bold {susy}}(\th-\th')\
\eeqa
where we used the  notations ${\one M}=M\otimes I$,
${\two M}=I \otimes M$. If $N=2$ supersymmetry is preserved, we write it
in terms of the solution (\ref{def}) as
\beqa
R_{\bold {susy}}(\theta)=K_{\bold {susy}}(\theta)Y_{\bold {susy}}(\theta)\
,\label{rsusy}
\eeqa
where the scalar factor $Y_{\bold {susy}}(\theta)$ has to be determined using some
physical assumptions (see below). Recalling the definition of the entries
(\ref{def}) it leads to
fourteen functional equations \cite{BK}:
\bea
 (i)&&a_-c_+
 \left(B D' -B' D \right)+a_-a_+
 [A ,A']=0\ ,\\
(ii) &&b_-b_+
 [A, E']+
 c_-c_+
 [E ,E']+\ c_-a_+
 \big(D B' -D' B \big)=0\ ,\\
(iii)&&c_-b_+
 \big(E A'-E'A \big)+ b_-c_+
 \big(A A'-E'E\big)+\ b_-a_+
 [B,D']=0\ ,\\
(iv)&&b_-b_+AD'  +
 c_-c_+ED'+\ c_-a_+DA'
 -\ a_-a_+D' A -\ a_-c_+E'D=0\ ,\\
 (v)&&b_-b_+B'A+\ c_-c_+ B'E+\ c_-a_+A'B-\ a_-a_+A B'-\ a_-c_+BE'= 0\ ,\\
(vi)&&b_-b_+D'E+\ c_-c_+D'A+\ c_-a_+E'D-\ a_-a_+ED'-\ a_-c_+DA'=0\ ,\\
(vii)&&b_-b_+EB'+\ c_-c_+AB'+\ c_-a_+BE'-\ a_-a_+B'E-\ a_-c_+A'B= 0\ ,\\
(viii)&&b_-a_+BE' +
 \ c_-b_+EB'+\ b_-c_+AB'
 -\ a_-b_+E'B=0\ ,\\
(ix)&&b_-a_+A'B+\ c_-b_+B'A+\ b_-c_+B'E-\ a_-b_+BA'=0\ ,\\
(x)&&b_-a_+E'D+\ c_-b_+D'E+\ b_-c_+D'A-\ a_-b_+DE'=0\ ,\\
(xi)&&b_-a_+DA'+\ c_-b_+AD'+\ b_-c_+ED'-\ a_-b_+A'D=0\ , \eea
where we used the shorthand notations $a_-=a(\theta-\theta')$,
$a_+=a(\theta+\theta')$ and similarly for $b$ and $c$ as well as
$A=A(\theta)$ and $A'=A(\theta')$ and similarly for $B, D$ and
$E$. The remaining three equations are obtained from $(i)$,
$(ii)$, $(iii)$ through the substitutions $A\leftrightarrow E$ and
$B\leftrightarrow D$. Straightforward  calculations show that
$K_{\bold {susy}}(\theta)$ given by (\ref{def}) with (\ref{entrees}) indeed satisfies all reflection equations above.

In \cite{GZ}, Ghoshal and Zamolodchikov proposed the use of the ``boundary unitarity'' and ``boundary cross-unitarity''
conditions to determine the overall factor $Y(\theta)$ associated with a
non-dynamical boundary. In case of a dynamical boundary, it can be applied directly \cite{BK} and gives the following
equations:
\beqa
&&\mbox{$\bullet$ \ Boundary unitarity:} \ \ \quad \qquad \qquad {R_{\bold {susy}}}^a_c(\theta)
{R_{\bold {susy}}}^c_b(-\theta)=
\delta^a_b{\mathbb I} \ \label{unit};\\
&&\mbox{$\bullet$ \ Boundary cross-unitarity:} \quad \qquad {R_{\bold {susy}}}^b_{\overline a}
(i\pi/2-\theta)=
{S_{\bold {susy}}}^{ab}_{a'b'}(2\theta){R_{\bold {susy}}}^{a'}_{\overline
b'}(i\pi/2+\theta)\label{crossunit}\ ,
\eeqa
where the
operator $\mathbb{I}$ denotes the identity which acts trivially on
the boundary ground state. We refer the reader to \cite{GZ} for details.
Let us now introduce two meromorphic functions \ $Y^{\bold{susy}}_{0}(\theta)$ \ and \ $Y^{\bold{susy}}_{1}(\theta)$
such that the prefactor in (\ref{rsusy}) is written
\beqa
Y_{\bold{susy}}(\theta)=Y^{\bold{susy}}_{0}(\theta)Y^{\bold{susy}}_{1}(\theta)\ \label{preftot}.
\eeqa
Using the explicit expressions  (\ref{entrees}),  they are chosen such that they solve the functional equations
\beqa &&Y^{\bold{susy}}_0(\theta)Y^{\bold{susy}}_0(-\theta)\ =\ 1 \ , \\
&&Y^{\bold{susy}}_0(i\pi/2-\theta)\ =\ \cos(u) Z(2u)
\ Y^{\bold{susy}}_0(i\pi/2+\theta)\ ,\ \nonumber \\
&&Y^{\bold{susy}}_1(\theta)Y^{\bold{susy}}_1(-\theta)\ =
\ \big[ -\sin^2(u/2)\cos^2(u/2) - \frac{1}{k^2}\sin^2(u/2) + \frac{1-\alpha^2+2k^2}{4k^4}\big]^{-1}\ ,\nonumber \\
&& Y^{\bold{susy}}_1(i\pi/2-\theta)\ =\ Y^{\bold{susy}}_1(i\pi/2+\theta)\
\nonumber\eeqa
in order to have (\ref{unit}) and (\ref{crossunit}). Using the results of \cite{GZ} and \cite{BK} for
 $\lambda=2/\betah^2-1$  we finally obtain
\beqa Y^{\bold{susy}}_0(\theta)=R_0(u)G_0(u)|_{\lambda=1/2}\label{refmat}\eeqa
where we used \cite{GZ}
\bea R_0(u)=\prod_{k=1}^{\infty}\Big[\frac{\Gamma(4\lambda k
-2\lambda u/\pi)\Gamma(1+4\lambda(k-1) -2\lambda
u/\pi)}{\Gamma(\lambda(4k-3) -2\lambda
u/\pi)\Gamma(1+\lambda(4k-1) -2\lambda
u/\pi)}/(u\rightarrow-u)\Big]\eea
and \cite{BK}
\beqa G_0(u)=
\prod_{k=1}^{\infty}\Big[\frac{\Gamma(1+(4k-2)\lambda-2\lambda
u/\pi)\Gamma((4k-2)\lambda-2\lambda
u/\pi)}{\Gamma((4k-4)\lambda-2\lambda
u/\pi)\Gamma(1+4k\lambda-2\lambda u/\pi)}/(u\rightarrow-u)\Big]\ .
\ \nonumber\eeqa
The other part is given by
\beqa Y^{\bold{susy}}_1(\theta)=\frac{\sigma(\eta,u)\sigma(i\vartheta,u)}
{\cos(\eta)\cosh(\vartheta)}|_{\lambda=1/2}\eeqa
with \cite{GZ}
\bea \sigma(x,v) = \frac{\cos x}{\cos(x+\lambda
v)}\prod_{l=1}^{\infty}\Big[\frac{\Gamma(1/2+(2l-1)\lambda+x/\pi-\lambda
v/\pi)\Gamma(1/2+(2l-1)\lambda-x/\pi-\lambda
v/\pi)}{\Gamma(1/2+(2l-2)\lambda-x/\pi-\lambda
v/\pi)\Gamma(1/2+2l\lambda+x/\pi-\lambda
v/\pi)}/(v\rightarrow-v)\Big] \ .\eea
Here \ $\eta$ \ and \ $\vartheta$ \ are two real IR boundary parameters related
with \ $k$ \ and the parameter \ $\xi$ \  by
\beqa
\cos(\eta)\cosh(\vartheta)=-\frac{1}{k}\cos{\xi} \qquad \ \ \mbox{and}\ \ \qquad
\cos^2(\eta)+\cosh^2(\vartheta)=1+\frac{1}{k^2}\ .\label{relpam}
\eeqa
Using these definitions, the parameter \ $\alpha$ \ entering in the boundary reflection  matrix  (\ref{rsusy})
with (\ref{def}) and (\ref{preftot}) becomes
\beqa
\alpha_\pm \ =\ \pm \sqrt {1 \ - \  2k^2\cos(2\xi)} \ .\label{defalpha}
\eeqa
It should be stressed that appart from the commutative case $\nu^{\bold{ir}}_{\pm}\equiv e^{\pm i\xi}$, there are no
values of the parameters for which the boundary reflection matrix (\ref{rsusy})  reduces to the Ghoshal-Zamolodchikov
one. Furthermore, we also checked that our solution can not be obtained from the Ghoshal-Zamolodchikov one using the
 fusion procedure suggested in \cite{Warn}. Notice that for the commutative case the exact relations between the IR
 boundary parameters \ $\{\eta,\vartheta\}$ \ and
the UV boundary parameters have been obtained by Al.B. Zamolodchikov \cite{Zamunpublished} and are supported by truncated
conformal space analysis \cite{Takacs}.

\subsection{The massive $N=2$ boundary superalgebra and boundary free energy}
Let us now see in which manner the presence of boundary fermionic degrees of freedom affects the $N=2$ supersymmetry
algebra. To see this, one can use the
one-particle asymptotic states $|u(\theta)\rangle,|d(\theta)\rangle$ representation (\ref{rep}) \cite{FendleyInt} as
before. They provide a representation  for the $N=2$ massive superalgebra \cite{WittenOlive}
\beqa &&\{{\cal Q}_+,{\cal Q}_-\}=2(H+P)\ ,\qquad \{{\overline
{\cal Q}}_+,{\overline {\cal Q}}_-\}=2(H-P) \ , \qquad \{{{\cal
Q}}_+,{\overline {\cal Q}}_-\}=2M{\cal N}\ , \qquad \{{\overline
{\cal Q}}_+,{ {\cal Q}}_-\}=2M{\cal N}\ ,
 \nonumber \\
&&{{\cal Q}}_\pm^2={\overline {\cal Q}}_\pm^2=0\ \qquad \mbox{and}
 \qquad \{{{\cal Q}}_\pm,F\}=\{{\overline {\cal Q}}_\pm,F\}=0 \ .
\eeqa
Using these relations, the expressions for the boundary supercharges (\ref{susycharges}) and anticommutation relations for the asymptotic boundary degrees of freedom (\ref{fermiondef}), it is straightforward to derive the relations
\beqa {\hat {\cal Q}}_\pm^2=2M{\cal N} \qquad \quad
\mbox{and}\qquad \quad \{{\hat {\cal Q}}_+,{\hat {\cal
Q}}_-\}=4\big(H -E^{\lambda=1/2}_{boundary}(-1)^{2F}\big)
 \ \qquad \mbox{where} \qquad E^{\lambda=1/2}_{boundary}\equiv\frac{M}{k^2} \ . \label{subalg}
\eeqa
Here it is interesting to notice that the term associated with the boundary energy is nothing but the boundary energy of the non-dynamical boundary SG model\,\footnote{Whereas we checked explicitly that it doesn't coincide with the boundary free energy calculated for general values of the coupling, i.e. in case of boundary degrees of freedom of the form (\ref{bop}).} at special point $\betah^2=4/3$, given by \cite{Zamunpublished} (see \cite{Takacs} for details)
\beqa
E_{boundary}^{\lambda}= -\frac{M}{2\cos(\pi/2\lambda)}\big[\cos(\eta/\lambda)+\cos(\vartheta/\lambda)-\frac{1}{2}\cos(\pi/2\lambda)+\frac{1}{2}\sin(\pi/2\lambda)-\frac{1}{2}\big] \
\eeqa
together with (\ref{relpam}).
Consequently, both terms are positive definite for any value of the parameter $k$ and any boundary ground state, as required. As an example, we can consider the massless limit \ $M\rightarrow 0$ \ with $k$ finite. In this case, the expectation value in the second term of (\ref{subalg})
takes its minimal value, independantly of $k$.
Notice that this subalgebra is different from the one given in \cite{Nepo2}. Although the second anticommutator is identical, the first differs by boundary contributions. Indeed, in \cite{Nepo2} a term proportional to the operator $F^2$ arises in the r.h.s. of the first anticommutator. In other words (appart from the massless case), in order to avoid negative (or complex) expectation values
of \ ${\hat {\cal Q}}_\pm^2$ \ Bogolmolnyi bounds are required in \cite{Nepo2} for the boundary parameters \ $(k,\xi)$. It is however not the case here.

\section{Comments about the boundary $N=2$ supersymmetric sine-Gordon}
In the bulk, the boundary $N=2$ supersymmetric sine-Gordon model possesses local (nonlocal) conserved charges which commute with each other \cite{Uematsu}. Together, they generate the quantum affine algebra $U_q({\widehat {sl_2}})\otimes U_{q^2=-1}({\widehat {sl_2}})$. It follows that the soliton $S$-matrix has the factorized form \cite{Uematsu}:
\beqa
S^{N=2}_{SG}(\theta)=S_{SG,\betah^2}(\theta)\otimes S_{SG,q^2=-1}(\theta)
\eeqa
where the bosonic part $S_{SG,\betah^2}(\theta)$ is the sine-Gordon scattering matrix. From the tensor product
structure, in the boundary case the boundary reflection matrix also takes a factorized form:
\beqa
R^{N=2}_{SG}(\theta)=R_{SG,\betah^2}(\theta)\otimes R_{\bold {susy}}(\theta)\ .\label{susySG}
\eeqa
In \cite{Nepo1}, an action for the $N=2$ supersymmetric sine-Gordon model has been proposed. It is exact in
 case of massless bulk and an approximation to first order in the bulk mass. At the boundary, this action
 contains fermionic boundary degrees of freedom coupled with the fermionic and bosonic fields
 $\psi^{\pm},{\overline \psi}^{\pm},\varphi^{\pm}$. Starting from this Lagrangean, we checked explicitly
  that the boundary supercharges take a form  similar to (\ref{susycharges}) with (\ref{opcond}) and (\ref{link}).
  They are different from the ones proposed in \cite{Nepo2}, which do not contain fermionic boundary degrees of freedom
  in front of the term associated with the topological charge.
Higher order corrections in the bulk mass term, if needed, would lead to the same form. Consequently, an exact action
for the $N=2$ supersymmetric boundary sine-Gordon model would lead to conserved boundary supercharges of the form
(\ref{susycharges}) generating the $N=2$ massive boundary superalgebra (\ref{subalg}), instead of the ones proposed
in \cite{Nepo2}. Using previous analysis, it follows that the supersymmetric part $R_{\bold {susy}}(\theta)$ of the
 reflection matrix (\ref{susySG}) is given by (\ref{rsusy}).

Let us now turn to the bosonic part. Actually, there are two kinds of Lagrangean that can be constructed. On one hand,
 with a boundary interaction of the form \cite{Nepo1} which only contains fermionic boundary degrees of freedom
 (here denoted ${\bold b}(y),{\bold b}^\dagger(y)$). In this case, the pure bosonic boundary reflection scattering
 matrix $R_{SG,\betah^2}(\theta)$ contains two free parameters and follows from the one proposed by Ghoshal-Zamolodchikov
 \cite{GZ}. On the other hand, it is well expected that a boundary interaction including {\it bosonic}
 (denoted ${\bold p}(y),{\bold q}(y)$) and fermionic boundary degrees of freedom can be constructed as well.
  In such case, the bosonic
boundary reflection matrix follows from the results of \cite{BK}, i.e. it reads
\beqa
R_{SG,\betah^2}(\theta)=K_0(\theta)Y(\theta)\
\label{ksg}
\eeqa
where $Y(\theta)$ ensures boundary unitarity and boundary cross-unitarity symmetry. Its ``minimal'' part $K_0(\theta)$
 takes the same form as in (\ref{def}) with entries
\beqa
&&A(\theta)=\pm\big(q^{-1}e^{\theta/\betah^2}\cosh({\bold p})-qe^{-\theta/\betah^2}\cosh({\bold q})\big)(q-q^{-1})/2c\ ,\nonumber \\
&&E(\theta)=\pm\big(q^{-1}e^{\theta/\betah^2}\cosh({\bold q})-qe^{-\theta/\betah^2}\cosh({\bold p})\big)(q-q^{-1})/2c\ ,\nonumber \\
&&B(\theta)=\Big(-c^2q^{-1}e^{2\theta/\betah^2}-c^2qe^{-2\theta/\betah^2}+\frac{q-q^{-1}}{q+q^{-1}}(q^{-1}\cosh({\bold q})\cosh({\bold p})
 - q\cosh({\bold p})\cosh({\bold q}))\Big)/2c^2\ ,\nonumber\\
&&D(\theta)=\Big(-c^2q^{-1}e^{2\theta/\betah^2}-c^2q
e^{-2\theta/\betah^2}+\frac{q-q^{-1}}{q+q^{-1}}(-q\cosh({\bold q})\cosh({\bold p})
 + q^{-1}\cosh({\bold p})\cosh({\bold q}))\Big)/2c^2\ .\label{entreessusy}
\eeqa
Here, the deformation parameter $q=
e^{i\pi/\betah^2}$ and $c=\sin(\pi/\betah^2)$.
Depending on the sign in (\ref{entreessusy}), the boundary quantization condition is fixed to
\beqa
\big[{\bold p},{\bold q}\big]=-\frac{2i\pi}{\betah^2}\ mod\ (2i \pi)\ .
\eeqa
In total, the boundary reflection matrix (\ref{susySG}) of each
model contains four or two boundary parameters (up to a canonical
transformation of the boundary degrees of freedom ${\bold p}$ and
${\bold q})$, respectively. Probably there exists some relation
between them, following the analysis of pole structure in
 \cite{Nepo2}. Details as well as further study of the second model will be considered elsewhere.

To conclude, notice that considering the analytic continuation $\betah^2=-b^2$ would provide  the factorized scattering
 theory for the (second) $N=2$ supersymmetric boundary sinh-Gordon model. In this case, up to the sign, the boundary
 quantization condition for boundary bosonic degrees of freedom becomes $\big[{\bold p},{\bold q}\big]=2i\pi/b^2$. Due
  to the recent conjecture \cite{stani} about the dual representation of $N=2$ supersymmetric Liouville field theory,
   it is worth interesting to understand weither a dual boundary action with  $\big[{\bold p},{\bold q}\big]=2i\pi b^2$
    can be explicitly constructed.

\section{Relations with quantum impurity problems}
Several quantum impurity problems can be analysed using analytical
methods. For strongly interacting systems, nonperturbative ones are
crucial in order to study bosonized versions of gapless (critical)
quantum systems (for a review, see for instance \cite{Saleur}). Among
the boundary quantum field theories that can describe such systems, the boundary massless SG model has been proposed and some exact results obtained from its nonperturbative analysis \cite{masslessSG}. Also, massive theories have applications in 1D impurity systems with an excitation gap. Let's see how our model provides exact results for such systems.\vspace{0.5cm}

$\bullet$ {\bf The boundary scaling Lee-Yang model}\\
The scaling limit of the Ising model with a purely (bulk) imaginary magnetic field is described by the Lee-Yang model. Its UV limit leads to the non-unitary ${\cal M}_{2/5}$ minimal conformal field theory (with central charge $c=-22/5$) which possesses only one primary field with conformal dimension $\Delta=-1/5$. The bulk perturbation is identified with this field. In the IR limit, there is only one specie of particles associated with the Faddeev-Zamolodchikov creation operator ``${\cal A}(\theta)$'' whose scattering is described by the bulk $S$-matrix \cite{MussCar}
\beqa
{\cal A}(\theta_1){\cal A}(\theta_2)=S(\theta_1-\theta_2){\cal A}(\theta_2){\cal A}(\theta_1)\qquad \quad\mbox{with}\qquad \quad S(\theta)=-(1/3)(2/3)\label{bLY}
\eeqa
and where we used the notations of the previous sections. In particular, the Lee-Yang model can be thought as the SG model at coupling constant $\betah^2=4/5$ $(\lambda=3/2)$. Then , the bulk $S$-matrix  factor (\ref{bLY}) can be obtained from the first SG breather calculated in \cite{ZZ} after projecting out the soliton sector.

On the half-line, there is a single relevant boundary perturbation of the BCFT with certain conformal boundary conditions \cite{Dorey}. The scattering of the fundamental particle on the boundary (with creation operator $B$) is described by the reflection factor $R(\theta)$ which satisfies
\beqa
R(\theta)R(-\theta)=1 \ ,\qquad R(i\pi/2-\theta)= S(2\theta)R(i\pi/2+\theta) \ ,\qquad \mbox{and} \qquad R(\theta)= S(2\theta)R(\theta+i\pi/3)R(\theta-i\pi/3)\ .\label{rLY}
\eeqa
Without degrees of freedom at the boundary, the boundary scaling Lee-Yang model has been studied in details  in \cite{Dorey} from both analytical and numerical approach. In case of boundary degrees of freedom, it is thus natural to expect some changes in the scattering data. Taking $\betah^2=4/5$ $(\lambda=3/2)$ in the model (\ref{actionSG}), it is not difficult to show that the corresponding minimal part of the soliton/antisoliton reflection matrix can be written in terms of (\ref{def}) with (\ref{entrees}) as\,\footnote{This solution is obvious as the boundary Yang-Baxter equations are invariant under the simultaneous change of sign in front of $B(\theta)$ and $D(\theta)$.}
\beqa
K(\theta)|_{\lambda=3/2}=\sigma_3K_{{\bold {susy}}}(3\theta)\sigma_3\qquad \quad \mbox{where}\qquad \quad \sigma_3=diag(1,-1)\ .
\eeqa
For \ $\lambda=3/2$ \ there is only one breather \ $n=1$. Using the boundary  bootstrap equation \cite{Fri95,Gho94}, it is straightforward to obtain the first breather reflection amplitude
\beqa R_{B}^{(1)}(\theta)=- R_0^{(1)}(u)
S^{(1)}(0,u)S^{(1)}(\pi/2,u)S^{(1)}(\eta,u)
S^{(1)}(i\vartheta,u)|_{\lambda=3/2}\ \label{R1B}\eeqa
where \cite{Gho94}
\beqa
R_0^{(n)}(u)=\frac{\big(\frac{1}{2}\big)\big(1+\frac{n}{2\lambda}\big)}
{\big(\frac{3}{2}+\frac{n}{2\lambda}\big)}\prod_{l=1}^{n-1}\frac{\big(\frac{l}{2\lambda}\big)\big(1+\frac{l}{2\lambda}\big)}
{\big(\frac{3}{2}+\frac{l}{2\lambda}\big)^2}\qquad \mbox{and}
\qquad
S^{(n)}(x,u)=\prod_{l=0}^{n-1}\frac{\big(\frac{x}{\lambda\pi}-\frac{1}{2}+\frac{n-2l-1}{2\lambda}\big)}
{\big(\frac{x}{\lambda\pi}+\frac{1}{2}+\frac{n-2l-1}{2\lambda}\big)} \ \label{defo}
\eeqa
with the shorthand notation \ $\big(x\big)=
\sin(u/2+x\pi/2)/\sin(u/2-x\pi/2)$. However,  in order to be a solution of (\ref{rLY}) it is necessary to restrict the values of the boundary parameters to \ $2\eta/\pi=\pm(b/2+2)$, \ $2\vartheta/\pi=\pm(b/2+1)$. The final result for the reflection factor reads
\beqa
R(\theta) = \Big(\frac{3}{6}\Big)^{-1}\Big(\frac{5}{6}\Big)^{-1}\Big(\frac{4}{6}\Big)^{-1}
\Big(\frac{1+b}{6}\Big)\Big(\frac{1-b}{6}\Big)^{-1}\Big(\frac{5+b}{6}\Big)^{-1}\Big(\frac{5-b}{6}\Big)\ .
\eeqa
Following the notations of \cite{Dorey}, for \ $b=0,1$ \ and \ $2$ \ one recovers the minimal solutions \ $R_{(2)}$, $R_{(4)}$ \ and \ $R_{(1)}$ \ of \ (\ref{rLY}), respectively. In particular, in \cite{Dorey} the boundary condition \ $\mathbb I$ \ was identified with the reflection factor \ $R_{(1)}$. In case of fermionic boundary degrees of freedom added, together with
\beqa
R(i\pi/2-\theta)=\frac{S(\theta-i(b-2)/6)}{S(\theta+i(b-2)/6)}R(i\pi/2+\theta)  \ \label{symm}
\eeqa
 we conclude that the fixed point $b=2$ of the symmetry transformation (\ref{symm}) is associated with the conformal boundary condition \ $\mathbb I$. Without fermionic boundary degrees of freedom, it was not the case in \cite{Dorey}. This phenomena clearly needs further investigation which however goes beyond the scope of this paper.\\

 $\bullet$ {\bf The anisotropic Kondo model and its massive extension}\\
 In \cite{Bass99} two integrable massive versions of the anisotropic spin $1/2$ Kondo model have been proposed at the reflectionless points. For general values of the coupling $\betah$, the model (\ref{actionSG}) is an other one \cite{BK}, although boundary degrees of freedom do not belong anymore to $su_q(2)$. From the results of the previous section,  we can now consider the Hamiltonian
 \beqa H_{MK}=\frac{1}{4\pi g}\int_{-\infty}^0 dx \Big((\pi(x))^2
 +(\partial_x\phi(x))^2 +8\pi g\mu\cos(2\phi(x))\Big) - \mu_B
 \big(S_+e^{i\phi(0)} + S_-e^{-i\phi(0)}\big) \ \quad \mbox{for}\quad g=\frac{2}{2n+3}\ \label{mk}\eeqa
where $n\in{\mathbb N}$ and $S_{\pm}$ are Pauli matrices. In the limit $\mu=0$, it corresponds to the usual anisotropic (at zero voltage) Kondo model. Taking $\phi_0=0$ and $\nu^{\bold{uv}}_\pm(y)=S_\pm(y)$ in (\ref{pertboundary}), the transformation $\phi \leftrightarrow -\phi$ with $S_+ \leftrightarrow S_-$
leaves the Hamiltonian invariant. Then, the corresponding soliton/antisoliton reflection matrix follows from (\ref{rsusy}) setting $\alpha=0$ and $\theta \rightarrow (2n+1)\theta$ in (\ref{entrees}), i.e. it reads
\beqa
{\cal R}^{(n)}(\theta)=K_{\bold {susy}}((2n+1)\theta)|_{\alpha=0}\Big[R_0(u)G_0(u)\frac{\sigma(\eta,u)\sigma(i\vartheta,u)}
{\cos(\eta)\cosh(\vartheta)}\Big]_{\lambda=n+1/2}\ .
\eeqa
Notice that the massive deformation for $n=0$
in (\ref{mk}) preserves the known $N=2$ supersymmetry of the massless Kondo model at that point \cite{fendleydual}. Breathers boundary reflection amplitudes  are calculated directly and given by
\beqa R_{B}^{(n)}(\theta)=(-1)^n R_0^{(n)}(u)
S^{(n)}(0,u)S^{(n)}(\pi/2,u)S^{(n)}(\eta,u)
S^{(n)}(i\vartheta,u)|_{\lambda=1/2+n}\ \label{RnB}\eeqa
together with (\ref{defo}). It shows that the massive extension proposed
in \cite{Bass99} is {\it also} integrable at the special points $g=2/(2n+3)$.
If one considers a non-zero voltage, an interesting question would be to check weither the duality $g \leftrightarrow 1/g$ \cite{fa,fe} still exists at these points in the massive case, or not.\\

$\bullet$ {\bf The boundary critical Ashkin-Teller model}\\
In two dimensions, the Ashkin-Teller model \cite{AT} corresponds to two
Ising models coupled by a local four spin interaction. The action associated with the critical line of the Ashkin-Teller
(${\mathbb Z}_4$) model in the bulk corresponds to a conformal field theory with central
charge $c=1$ (a free massless scalar field $\phi$ \cite{Kadanoff1}). This critical
line can be parametrized by the conformal dimension
\ $\Delta_\epsilon=\betah^2/2$ \ of the thermal operator \
$\epsilon\equiv{\sqrt 2}\cos(\betah\phi)$. Then, its $\epsilon$-perturbation
 coincides with the
sine-Gordon model. The order
 parameters are the fields $\sigma$, $\sigma^\dagger$ and the field
 $\Sigma\sim\sigma^2$ with conformal dimensions $\Delta=1/16$
 (independently of $\betah $ \cite{Kadanoff2}) and  $\Delta_\Sigma=\betah^2/8$,
 respectively. The field $\Sigma$ which is local with respect to
 $\phi$ can be realized in terms of $\exp(\pm\betah\phi/2)$. At the special point $\betah^2=4/3$, the Ashkin-Teller
  model enjoys $N=2$ supersymmetry.

For generic values of the coupling $\betah$, a natural boundary
version of the Ashkin-Teller model can be associated with action
(\ref{actionSG}) and (\ref{bop}). At the special points
$\betah^2=4/(2n+3)$, it admits a two-parameter family of boundary
Lagrangean representation of the form
\beqa
{\cal A}_{bAT}=\int_{-\infty}^{\infty}dy\int_{-\infty}^{0}dx
\Big(\frac{1}{8\pi}(\partial_\nu\phi)^2-2\kappa\ \epsilon(x,y)\Big)
+ \int_{-\infty}^{\infty}dy\ \big(\nu_0 \ {\bold b}(y) + {\overline \nu}_0 \
{\bold b}^\dagger(y)\big)\ \Sigma(y) \ + \ {\cal A}_{boundary} \label{bAT}
\eeqa
where $\nu_0$ is a complex parameter  and
the boundary dynamics are  described by
\beqa
{\cal A}_{boundary}=i\int_{-\infty}^{\infty}{\bold b}^\dagger\frac{d}{dy}
{\bold b} \
dy\ .
\eeqa
The factorized scattering theory associated with these two phases follows from the results of \cite{BK} and those presented here. In particular, for $n=0$ this model exhibits $N=2$ boundary supersymmetry defined in section 2.
\vspace{0.5cm}

\noindent{\bf Acknowledgements:}  P.B. thanks P.E. Dorey, A. Furusaki, C. Mudry, H. Saleur, R. Sasaki, M. Stanishkov and R. Tateo for discussions and R.I. Nepomechie for comments. P.B's work supported by JSPS
fellowship.

\end{document}